\documentclass[aps,pre,twocolumn,groupedaddress]{revtex4-1}

\usepackage{graphicx} 
\usepackage{amsmath} 
\usepackage{amssymb}	
\newcommand{\dz}[0]{\Delta Z} 
\newcommand{\dphi}[0]{{\Delta \phi}}
\newcommand{\e}[0]{\epsilon} 
\newcommand{\eq}[1]{\begin{align} #1 \end{align}}

\newcommand{\avg}[1]{\left< #1 \right>} 
\newcommand{\driv}[2]{\frac{d #1}{d #2}} 
\newcommand{\dd}[2]{\frac{d^2 #1}{d #2^2}} 

\newcommand{\keff}{k_{\mathrm{eff}}}

\DeclareMathOperator{\Tr}{Tr}

\usepackage{hyperref} 
\hypersetup{
    colorlinks,
    citecolor=blue,
    filecolor=black,
    linkcolor=blue,
    urlcolor=blue
}

\begin{document}


\title{Scaling theory for the jamming transition}


\author{Carl P. Goodrich}
\email[]{goodrich@g.harvard.edu}
\altaffiliation[Now at the ]{School of Engineering and Applied Sciences, Harvard University, Cambridge, MA 02138, USA}

\author{Andrea J. Liu}
\email[]{ajliu@physics.upenn.edu}
\affiliation{Department of Physics, University of Pennsylvania, Philadelphia, Pennsylvania 19104, USA}

\author{James P. Sethna}
\email[]{sethna@lassp.cornell.edu}
\affiliation{Department of Physics, Cornell University, Ithaca, New York 14850, USA}

\date{\today}

\begin{abstract}
We propose a scaling ansatz for the elastic energy of a system near the critical jamming transition in terms of three relevant fields: the compressive strain $\dphi$ relative to the critical jammed state, the shear strain $\e$, and the inverse system size $1/N$.  We also use $\dz$, the number of contacts relative to the minimum required at jamming, as an underlying control parameter.  Our scaling theory predicts new exponents, exponent equalities and scaling collapses for energy, pressure and shear stress that we verify with numerical simulations of jammed packings of soft spheres.  It also yields new insight into why the shear and bulk moduli exhibit different scalings; the difference arises because the shear stress vanishes as $1/\sqrt{N}$ while the pressure approaches a constant in the thermodynamic limit.  The success of the scaling ansatz implies that the jamming transition exhibits an emergent scale invariance, and that it should be possible to develop a renormalization-group theory for jamming.

\end{abstract}


\maketitle

The existence of criticality at the jamming transition suggests that universal physics underlies rigidity in disordered solids ranging from glasses to granular materials~\cite{Liu:2010jx}.   The jamming transition marks the onset of rigidity in
athermal sphere packings, and was originally proposed as a zero-temperature transition~\cite{OHern:2002bs,OHern:2003vq} for soft repulsive spheres
in a non-equilibrium ``jamming phase diagram"~\cite{Liu:1998up} of varying packing density and applied shear.  Many studies have documented behaviors characteristic of critical phenomena near the jamming transition, including power-law scaling~\cite{Durian:1995eo,OHern:2002bs,OHern:2003vq} and scaling collapses~\cite{Ellenbroek:2006df,Olsson:2007df,Ellenbroek:2009dp,DagoisBohy:2012dh,Goodrich:2012ck,Tighe:2012gm,Goodrich:2014iu,vanDeen:2014kl} of numerous properties, diverging length scales~\cite{Silbert:2005vw,Wyart:2005wv,Ellenbroek:2006df,During:2012bsa,Goodrich:2013ke,Schoenholz:2013jv,Lerner:2014fn} and finite-size scaling~\cite{Goodrich:2012ck,Goodrich:2014iu,Vagberg:2011fe}. Theories have been developed to understand and relate some of these behaviors~\cite{Wyart:2005wv,Wyart:2005vu,During:2012bsa,Charbonneau:2014kk}, but a number of confusing issues have precluded a unified scaling analysis. Here we resolve these issues to present a renormalization-group-inspired scaling {\em ansatz} for the jamming critical point in terms of the fields originally identified by the jamming phase diagram, namely density and shear.

The critical-point scaling ansatz introduced by Widom~\cite{Widom:1965fl} 
in the 1960s set the stage for the development of the renormalization group.
The ansatz writes the free energy and correlation functions near continuous
phase transitions in terms of power-law ratios of the control parameters. 
Scaling predictions, e.g. for the magnetization, susceptibility, and specific
heat in magnetic systems, are derived from derivatives of the free
energy, thus providing a elegant and comprehensive description of
systems exhibiting what later was realized to be an emergent scale invariance.
By starting with a natural scaling ansatz for the elastic energy for a system just above the jamming transition, we show that
one can use the connectivity of the system to reveal robust and 
universal scaling functions for the energy, excess packing fraction,
shear strain, system size, pressure, shear stress, bulk modulus,
and shear modulus.

We consider disordered systems of $N$ soft spheres in a $d$-dimensional periodic box of volume $V$. Systems are at temperature $T=0$ and thus sit in a local minimum of the energy landscape, defined by the pairwise interaction potential
\eq{ U(r_{ij}) = \frac{U_0}{\alpha}\left(1-\frac{r_{ij}}{R_i+R_j}\right)^\alpha \Theta\left(1-\frac{r_{ij}}{R_i+R_j}\right), \label{eq:pairwise_interaction}	}
where $r_{ij}$ is the distance between the centers of particles $i$ and $j$, $R_i$ and $R_j$ are the particles' radii, $\Theta(x)$ is the Heaviside step function and $U_0$ sets the energy scale. The packing fraction is $\phi = \frac 1V \sum_i V_i$, where $V_i$ is the $d$-dimensional volume of particle $i$, and the shear strain $\e$ is defined relative to the strain of the as-quenched state. 
We define an effective spring constant, $\keff =U_0 (\alpha-1)/D_\text{avg}^2 \dz^{2(\alpha-2)}$~\cite{Vitelli:2010fa}, where $D_\text{avg}$ is the average particle diameter.  We then rescale energy, pressure, shear stress, bulk modulus and shear modulus by $\keff$ with appropriate factors of $D_\text{avg}$ so that $E=\textrm{energy}/\keff D_\text{avg}^2$, $p=\textrm{pressure}*D_\text{avg}/\keff$, and so on.  The scaling ansatz presented below in Eq.~\eqref{eq:scaling_ansatz} is for these scaled quantities, and does not depend on the exponent $\alpha$ in Eq.~\eqref{eq:pairwise_interaction}.

%
For the jammed systems considered here, the average number of interacting neighbors per particle (the contact number, $Z$) satisfies $Z \ge Z_\text{min}$, where $Z_\text{min} = 2d - (2d-2)/N$~\cite{OHern:2003vq,Goodrich:2012ck}; we define $\dz=Z-Z_\text{min}$.  Note that the packing fraction at the jamming transition, $\phi_{\text{c},\Lambda}$, varies from one
member of the ensemble $\Lambda$ to the next.  For each packing, we 
characterize the distance above the transition by $\Delta \phi \equiv \phi-\phi_{\text{c},\Lambda}$. 

This configuration-dependent critical density is a confusing feature of jamming that has impeded development of a scaling theory.
Analyzing the jamming transition in 
terms of the infinite-system critical density $\phi_{\text{c},\infty}$
leads to a quite different scaling picture, where the finite-size scaling 
behavior is dominated by the sample-to-sample fluctuations of
$\phi_{\text{c},\Lambda} - \phi_{\text{c},\infty}$~\cite{OHern:2002bs,OHern:2003vq,Vagberg:2011fy,Vagberg:2011fe,Liu:2014gu,Goodrich:2014iu}. 
This confusing behavior is shared with many other systems with sharp, global transitions in behavior, as originally discovered in the depinning of charge-density waves~\cite{MyersS93A,MyersS93B,Middleton:1993vc}.
Such systems may not obey the inequality between the correlation length and dimension $\nu \ge 2/d$ derived for equilibrium
systems, unless analyzed using deviations from the infinite system critical
point~\cite{PazmandiSZ97}. Here we use the system-dependent critical
density as suggested by Refs.~\cite{OHern:2002bs,OHern:2003vq}, which allows closer
scrutiny of scaling near the jamming transition.

For a given protocol for preparing jammed states, the mean energy density $E$ of a sphere packing will depend on $\dz$, $\dphi$,
$\e$ and $N$. Motivated by renormalization-group predictions for
systems with emergent scale invariance (discussed later), we
make the scaling ansatz 
\eq{	E\left(\dz,\dphi,\e,N\right) = \dz^\zeta \mathcal{E}_0\left(\frac{\dphi}{\dz^{\delta_{\dphi}}}, \frac{\e}{\dz^{\delta_\e}},  N \dz^\psi \right), \label{eq:scaling_ansatz}}
where the scaling exponents $\zeta$, $\delta_{\dphi}$, $\delta_\e$ and $\psi$ are yet to be determined.
Equation~\eqref{eq:scaling_ansatz} is set up so that the singular part of the energy in the thermodynamic limit is described in terms of $\dz$ and the scaling exponent $\zeta$, while the non-singular effects are described by the universal scaling function $\mathcal{E}_0$.  
The excess packing fraction and shear strain represent the two independent global deformations that are invariant to rotations and will be treated as distinct fields with scaling exponents $\delta_\dphi$ and $\delta_\e$, respectively.  The effects of finite system size
$N \sim L^d$, where $L$ is the (more traditional) system length, are included in the last argument. As jamming
is believed to have an upper critical dimension of
$d_u=2$~\cite{Wyart:2005jna,Wyart:2005vu,Liu:2010jx,Charbonneau:2011fg,Charbonneau:2012fl,Goodrich:2012ck,Charbonneau:2015ec},
finite-size effects should scale with $N\sim L^d$~\cite{BINDER:1985vl,Wittmann:2014fp}, so this choice should allow the critical
exponents to be independent of dimension for $d \ge 2$, with corrections to scaling likely in the upper critical dimension $d=2$.

Note that we scale the density shift and shear strain with different
critical exponents. This would seem natural, since they are different ``relevant
directions" in the jamming phase diagram. The different exponents lead to different scaling properties for the corresponding
susceptibilities -- the bulk and shear moduli, $B$ and $G$. Clearly, shear and bulk moduli
need not scale together -- liquids form a counterexample. The
critical jamming transition separating a non-equilibrium jammed
solid from a repulsive gas of non-overlapping spheres shares features of a liquid
($G=0$, $B>0$). This is reminiscent of the metal-like
resistivity at the two-dimensional disordered critical point separating
superconductors and insulators~\cite{Markovic:1998wz}.  

The different scalings of $B$ and $G$ give rise to different
scaling behavior in the transverse and longitudinal speeds of sound,
leading to distinct crossover length scales~\cite{Silbert:2005vw}. The existence of two distinct crossover lengths has been a second source of confusion impeding a unified understanding of scaling properties, since it has been unclear whether both scales should be incorporated into a finite-size scaling description.  Here we take advantage of earlier finite-size scaling collapses, which indicate that a single exponent controls finite-size scaling~\cite{Goodrich:2012ck}.  

Jammed packings at a given $\dphi$, $N$ and $\e$ have a prescribed average value of $\dz$.  
This has been a third source of confusion because, as we will show, it is useful to treat $\dz$ as an independent variable so that Eq.~\eqref{eq:scaling_ansatz} depends on four variables. The introduction of $\dz$ as an independent variable is 
analogous to the use of a variable magnetization in Landau theory of magnets, where the free energy density at fixed external
field and temperature is expressed as a function of magnetization, even though the equilibrium magnetization is set
by the field and temperature.

Finally, note that we shall quote integer and half-integer values for the various
critical exponents. This is motivated not only by the close agreement
with numerical simulations, but also by the belief that jamming is mean-field
in $d \ge 2$~\cite{Wyart:2005jna,Wyart:2005vu,Liu:2010jx,Charbonneau:2011fg,Charbonneau:2012fl,Goodrich:2012ck,Charbonneau:2015ec}. One should be
warned, however, that although mean-field theories often have critical
exponents as simple rational numbers, the current mean-field theories 
of jamming in hard-sphere packings have some irrational critical exponents~\cite{Charbonneau:2014kk,Yoshino:2014dz}. 

We proceed by calculating derivatives of the energy
(Eq.~\eqref{eq:scaling_ansatz}) with respect to the fields $\dphi$ and $\e$: 
\eq{ 
	p \equiv  \phi \driv E {\dphi} &= \dz^{\beta_p} \mathcal{P}_0\left( \frac{\dphi}{\dz^{\delta_{\dphi}}},  \frac{\e}{\dz^{\delta_\epsilon}} ,  N\dz^{\psi}    \right), \label{eq:p_scaling_form}\\
	s \equiv \driv E {\e} &= \dz^{\beta_s} \mathcal{S}_0\left( \frac{\dphi}{\dz^{\delta_{\dphi}}},  \frac{\e}{\dz^{\delta_\e}} ,  N\dz^{\psi}   \right). \label{eq:s_scaling_form}
}
Here we see that the pressure and shear stress arise naturally as order parameters that depend on powers of the excess contact number, $\dz$.  Until now, it has not been clear whether the natural control variable for the problem (analogous to reduced temperature in the Ising model) should be $\dz$, $\dphi$ or $p$.  This has been a fourth source of confusion preventing construction of a scaling theory.  In our formulation, it is clear that the proper control variable is $\dz$, and that $p$ and $\sigma$ are analogous to the magnetization while $\dphi$ and $\e$ are analogous to the magnetic field in the Ising model.

Similarly, the shear and bulk moduli are second derivatives of the energy and are interpreted as susceptibilities, taking the form
\eq{
	B\equiv  \frac{\phi^2}2 \dd{E}{\dphi} &= \dz^{\gamma_B} \mathcal{B}_0\left( \frac{\dphi}{\dz^{\delta_{\dphi}}},  \frac{\e}{\dz^{\delta_\e}} ,  N\dz^{\psi}    \right), \label{eq:B_scaling_form} \\
	G \equiv \dd{E}{\e} &= \dz^{\gamma_G}\mathcal{G}_0\left( \frac{\dphi}{\dz^{\delta_{\dphi}}},  \frac{\e}{\dz^{\delta_\e}} ,  N\dz^{\psi}    \right). \label{eq:G_scaling_form}
}
The calculation of these scaling forms imply four exponent relations:
\eq{
\beta_p &= \zeta - \delta_\dphi, ~~~~~~~~~ \beta_s = \zeta-\delta_\e \nonumber\\
\gamma_B &= \zeta-2\delta_\dphi, \mathrm{~and~} \gamma_G = \zeta-2\delta_\e.
\label{eq:four_exponent_relations}
}
Note that the factors of $\phi$ and $\phi^{2}$ that appear in Eqs.~\eqref{eq:p_scaling_form} and \eqref{eq:B_scaling_form} are slowly varying and can be treated as constant near the singularity.
Also note that the scaling functions $\mathcal{P}_0$, $\mathcal{S}_0$, $\mathcal{B}_0$, and $\mathcal{G}_0$ can be written explicitly as functions of $\mathcal{E}_0$ and its derivatives.

While we have written our scaling theory in terms of the excess packing fraction, in practice one commonly creates packings at fixed values of $p$ instead of $\Delta \phi$, resulting in a ``fixed $p \e N$" ensemble. Equations~\eqref{eq:scaling_ansatz}-\eqref{eq:G_scaling_form} can be easily adapted to this ensemble by inverting Eq.~\eqref{eq:p_scaling_form},
\eq{	\frac{\dphi}{\dz^{\delta_\dphi}} = \Phi\left( \frac{p}{\dz^{\beta_p}},  \frac{\e}{\dz^{\delta_\e}} ,  N\dz^{\psi}    \right), \label{eq:dphi_scaling_withp} }
and inserting this into Eqs.~\eqref{eq:scaling_ansatz} and \eqref{eq:s_scaling_form}-\eqref{eq:G_scaling_form}. The other scaling functions can then be written as functions of the scaled pressure, shear strain and system size, taking the form
\eq{ 
	E &= \dz^\zeta \mathcal{E}\left(\frac{p}{\dz^{\beta_p}}, \frac{\e}{\dz^{\delta_\e}}, N \dz^\psi \right), \label{eq:E_scaling_withp}\\
	s &= \dz^{\beta_s} \mathcal{S}\left(\frac{p}{\dz^{\beta_p}}, \frac{\e}{\dz^{\delta_\e}}, N \dz^\psi \right), \label{eq:s_scaling_withp} \\
	B &= \dz^{\gamma_B} \mathcal{B}\left(\frac{p}{\dz^{\beta_p}}, \frac{\e}{\dz^{\delta_\e}}, N \dz^\psi \right),\\
	G &= \dz^{\gamma_G}\mathcal{G}\left(\frac{p}{\dz^{\beta_p}}, \frac{\e}{\dz^{\delta_\e}}, N \dz^\psi \right). \label{eq:G_scaling_withp}
}
Note that the scaling functions in Eqs.~\eqref{eq:E_scaling_withp}-\eqref{eq:G_scaling_withp} are different than those appearing in Eqs.~\eqref{eq:scaling_ansatz}-\eqref{eq:G_scaling_form}. It is straightforward to adjust the theory to other ensembles, such as the fixed-$\dphi \;\! s N$ ensemble of the original jamming phase diagram~\cite{Liu:1998up}
, or the fixed-$p s N$ ensemble of Ref.~\cite{DagoisBohy:2012dh}.

Recall that $\dz$ is not externally controlled but instead is measured for packings at a given $p$, $\e$ and $N$, forming the probability distribution $R(\dz | p, \e, N)$. 
 Starting from the scaling hypothesis, Appendix~\ref{sec:dz_distribution} shows that the relation between the average of $\dz$ and $p$, $\e$ and $N$ takes the form
\eq{	
	\frac{p}{\dz^{\beta_p}} &= F_2\left(\frac{\e}{\dz^{\delta_\e}}, N \dz^\psi \right),
	\label{eq:p_scaling_2}
}
where the scaling function $F_2$ could depend on the details of the numerical protocol used to create the systems. 
Due to this interdependency between the scaling variables, we see that the scaling functions in Eqs.~\eqref{eq:dphi_scaling_withp}-\eqref{eq:G_scaling_withp} can be written in terms of just two variables, $\e/\dz^{\delta_\e}$ and $N \dz^\psi$. Furthermore, for unsheared packings, $\e=0$ by our definition (see the discussion following Eq.~\eqref{eq:pairwise_interaction}),
 so in that case the number of variables in the scaling functions is reduced to only one, $N \dz^\psi$. Therefore, our theory predicts
\eq{ 
	E &= \dz^\zeta \mathcal{E}\left( N \dz^\psi \right), \label{eq:E_scaling_numerical}\\
	\dphi &= \dz^{\delta_\dphi}\Phi\left(   N\dz^{\psi}    \right), \label{eq:dphi_scaling_numerical} \\
	s &= \dz^{\beta_s} \mathcal{S}\left(N \dz^\psi \right), \label{eq:s_scaling_numerical}\\
	B &= \dz^{\gamma_B} \mathcal{B}\left( N \dz^\psi \right), \label{eq:B_scaling_numerical}\\
	G &= \dz^{\gamma_G}\mathcal{G}\left(N \dz^\psi \right), \label{eq:G_scaling_numerical}
}
as well as (see Appendix~\ref{sec:dz_distribution})
\eq{	N^{1/\psi} \dz= \mathcal{Z}\left(Np^{\psi/\beta_p} \right). \label{eq:avg_dz_p}	}

\begin{table}
\begin{tabular*}{0.45\textwidth}{@{\extracolsep{\fill}} |c|cccc@{\ \ \ }|cccc|}
	\hline
	exponent & $\zeta$ & $\delta_\dphi$ & $\delta_\e$ & $\psi$ & $\beta_p$ & $\beta_s$ & $\gamma_B$ & $\gamma_G$ \\ 
	\hline
	value & $4$ & $2$ & $3/2$ & $1$\ & $2$ & $5/2$ & $0$ & $1$ \\ 
	\hline
\end{tabular*}
\caption{\label{table:list_of_exponents}List of scaling exponents and their
approximate values. As discussed in the text, numerical studies 
suggest that jamming exponents are close to the integer or half-integer 
values present here. These eight exponents are related by five known exponent
relations (Eqs.~\eqref{eq:four_exponent_relations} and~\eqref{eq:PvsS_exponent_relation}), leaving three independent critical exponents.}
\end{table}

\begin{figure}
	\centering
	\includegraphics[width=\linewidth]{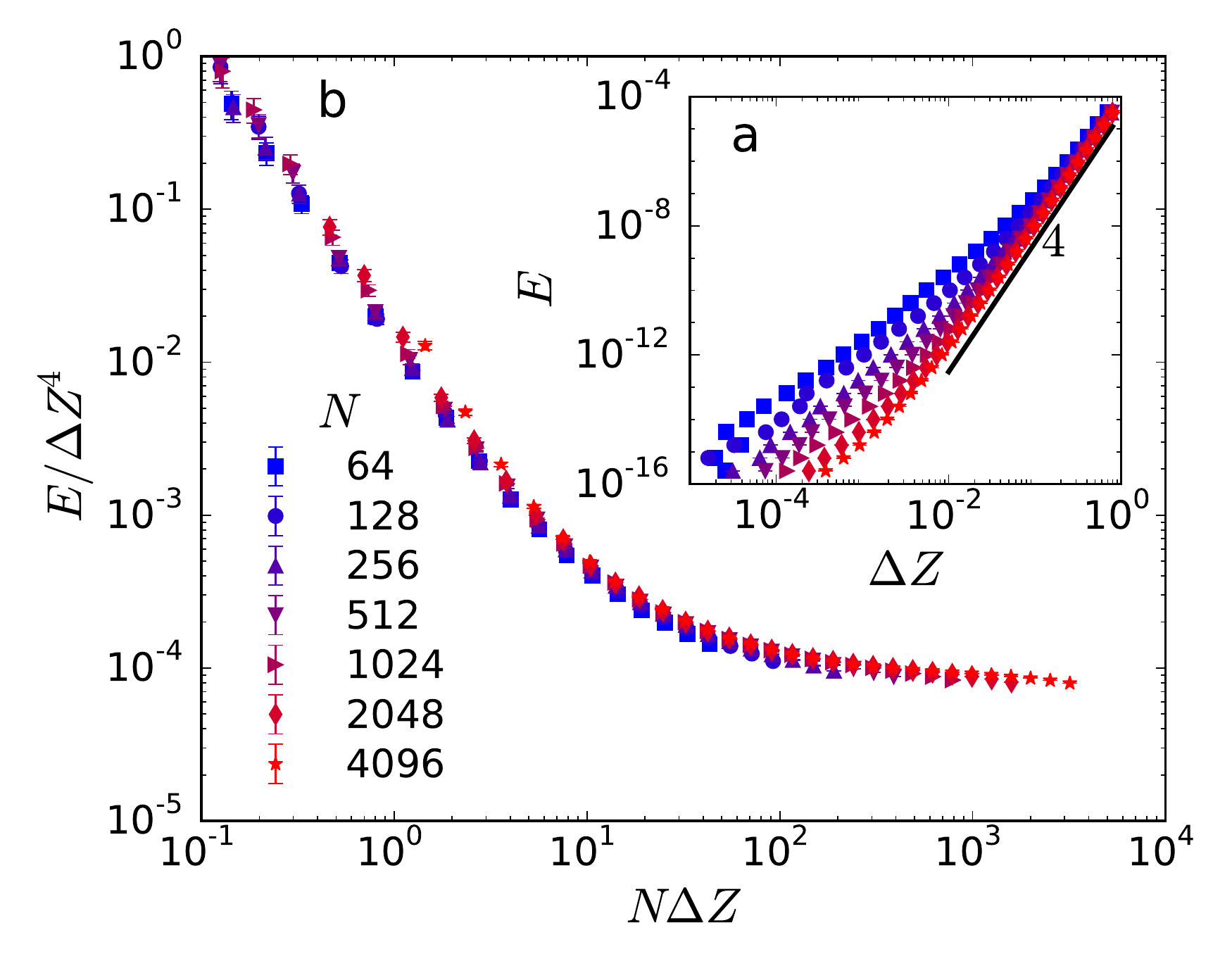}
	\caption{\label{fig:energy_scaling}Scaling collapse of the energy. a) Energy as a function of $\dz$. b) Energy scaled according to Eq.~\eqref{eq:scaling_ansatz} using the predicted exponents $\zeta = 4$ and $\psi=1$.} 
\end{figure} 

\begin{figure}
	\centering
	\includegraphics[width=\linewidth]{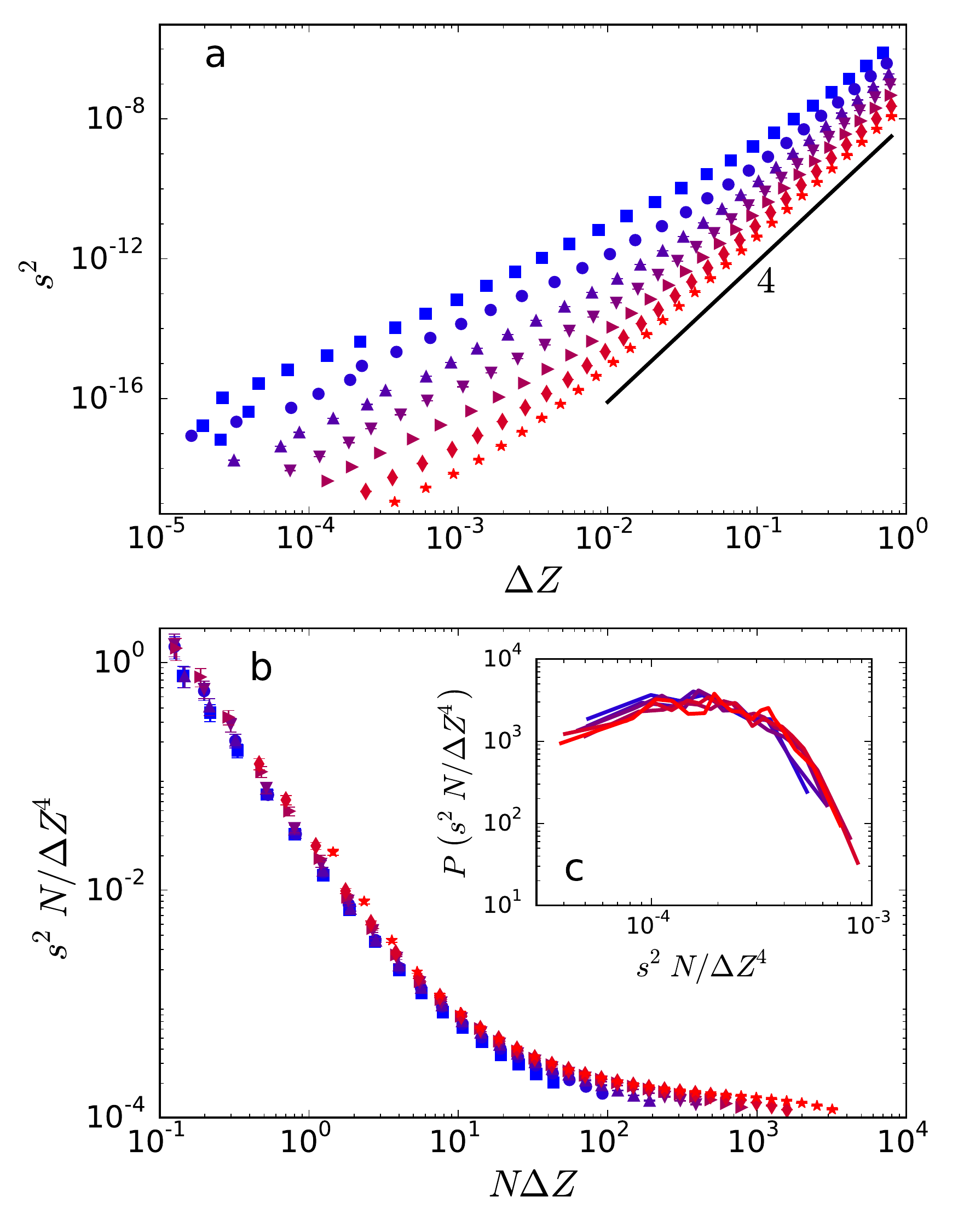}
	\caption{\label{fig:s2_scaling}Scaling collapse of the residual shear stress. a) $s^2$ as a function of $\dz$. b) $s^2$ scaled according to Eq.~\eqref{eq:s_scaling_numerical2} using the predicted exponents $\beta_s = 5/2$ and $\psi=1$. 
	c) Probability distribution of $s^2N/\dz^4$ for systems at fixed $N\dz$ ($50 \leq N\dz \leq 52$) and fixed $p/\dz^2$ ($7.9\times 10^{-3} \leq p/\dz^2 \leq 8.3\times 10^{-3}$). For $N=64$, insufficient data exists in these ranges to calculate a distribution. Symbols and colors have the same meaning as in Fig.~\ref{fig:energy_scaling}.} 
\end{figure}

We now appeal to known numerical results for $\dz$, $B$ and $G$ in the $p\e N$ ensemble to obtain the scaling exponents.
Note that the functions $\mathcal{B}$, $\mathcal{G}$, and $\mathcal{Z}$ should become independent of $N$ in the large $N$ limit, and it is well established that $\dz \sim p^{1/2}$, $B\sim \dz^0$ and $G\sim \dz$ in this regime~\cite{Durian:1995eo,OHern:2003vq,Goodrich:2014iu}.
Furthermore, recent studies of finite-size effects found that $N \dz = F(N p^{1/2})$~\cite{Goodrich:2012ck,Goodrich:2014iu}\footnote{Note that the finite-size analysis of Refs.~\cite{Goodrich:2012ck,Goodrich:2014iu} also found that $GN = F_G\left(pN^2\right)$, which is consistent with Eq.~\eqref{eq:G_scaling_numerical} and the exponents in Table~\ref{table:list_of_exponents}, thus providing the first check of our scaling theory}.
Comparing these results to Eqs.~\eqref{eq:B_scaling_numerical}-\eqref{eq:avg_dz_p}, we see that $\beta_p = 2$, $\gamma_B = 0$, $\gamma_G = 1$ and $\psi=1$. Using the exponent equalities of Eq.~\eqref{eq:four_exponent_relations}, we also find $\zeta = 4$, $\delta_{\dphi} = 2$, $\delta_\e = 3/2$ and $\beta_s = 5/2$. These exponents are summarized in Table~\ref{table:list_of_exponents}.

Note that the scaling exponents can be obtained by appealing to theoretical arguments for the scaling of the shear modulus and contact number~\cite{Wyart:2005vu} and to finite-size results~\cite{Goodrich:2012ck} instead of purely to numerical results.   

Our scaling theory yields predictions for the scaling of $E$, $\dphi$ and $s$ in Eqs.~\eqref{eq:E_scaling_numerical}-\eqref{eq:s_scaling_numerical}. To test these predictions, we generate jammed sphere packings in the $p\e N$ ensemble, using the algorithm of Refs.~\cite{Goodrich:2012ck,Goodrich:2014iu} to produce mechanically stable packings over a range of pressures (between $p=10^{-8}$ and $p=10^{-1}$) and system sizes (between $N=64$ to $N=4096$) with $\e=0$. We focus here on $d=3$ systems with harmonic interactions ($\alpha=2$ in Eq.~\eqref{eq:pairwise_interaction}) and a 50-50 bidisperse mixture of particles with radius ratio 1:1.4. Two-dimensional data, shown in Appendix~\ref{sec:corrections_to_scaling}, exhibit corrections to scaling expected at the upper critical dimension but are otherwise consistent with our theory.

We first consider the energy density $E$, which is shown as a function of $\dz$ in Fig.~\ref{fig:energy_scaling}a. 
Each point is an average over approximately 5000 configurations at a given $p$ and $N$. The clear finite-size effects collapse quite well when the data are scaled according to Eq.~\eqref{eq:E_scaling_numerical} with $\zeta = 4$ and $\psi=1$ (see Fig.~\ref{fig:energy_scaling}b), as predicted.

We now turn to the scaling of shear stress $s$.  Our packings are prepared at $\e=0$ so the stress fluctuates around zero.  We therefore consider the average of $s^2 \equiv \Tr \left[\sigma - \tfrac 1d \Tr{\sigma}\right]^2$, where $\sigma$ is the stress tensor.  This average is shown as a function of $\dz$ in Fig.~\ref{fig:s2_scaling}a.  To collapse these data, we note that $s^2$ vanishes in the infinite system size limit as $1/N$ (see Appendix \ref{sec:sp_exponent}), consistent with the central limit theorem.  
It is therefore convenient to factor out one power of $\left(N\dz^\psi\right)^{-1}$ from the scaling function for $s^2$ analogous to Eq.~\eqref{eq:s_scaling_numerical}:
\eq{	s^2 = \dz^{2\beta_s} \left[ \left(N\dz^{\psi}\right)^{-1} \mathcal{S}_2\left(N\dz^\psi\right) \right].	\label{eq:s_scaling_numerical2}}
This predicted scaling collapse is verified numerically in Fig.~\ref{fig:s2_scaling}b with $\beta_s=5/2$ and $\psi=1$. Finally, as expected from classical scaling theories, the inset in Fig.~\ref{fig:s2_scaling} demonstrates that our theory also describes the distributions of quantities, in this case $s^2$.

The fact that the mean stress (unlike the pressure) remains zero for systems above the jamming transition, and that the variance of the stress fluctuations scale as $1/N$, explains why the bulk modulus scales differently from the shear modulus at the jamming transition. In Appendix~\ref{sec:sp_exponent} we analyze the pressure and stress fluctuations microscopically, and derive an important exponent relation, $2\beta_s-\psi =2 \beta_p$ (Eq.~\eqref{eq:PvsS_exponent_relation}), between the singularities of stress and pressure which also yields the relation $\gamma_G=\gamma_B-\psi$ between shear and bulk moduli. The analysis in Appendix~\ref{sec:sp_exponent} uses the lack of long-range bond orientational order to derive this exponent relation; hence we predict that shear-jammed systems~\cite{Bi:2011bt} will have some components of their shear modulus that scale as the bulk modulus as jamming is approached (since the shear jamming will yield long-ranged bond orientation correlations).

Note that Eqs.~\eqref{eq:p_scaling_form} and \eqref{eq:s_scaling_form} are stress-strain relations, and the well-known compressional stress relation $p \sim \dphi$~\cite{Durian:1995eo,OHern:2003vq} emerges from the scaling ansatz.  Similarly, the scaling collapse of the shear stress-strain relation, obtained by integrating Eq.~\eqref{eq:G_scaling_withp} as in Appendix \ref{sec:dz_distribution}, so that $s=\dz^{\beta_s} \mathcal{S_1}\left(\frac{\e}{\dz^{\delta_\e}},N \dz^\psi \right)$, is consistent with that obtained earlier for harmonic spring networks~\cite{Wyart:2008jg}. The scaling ansatz also yields a prediction for the scaling of the excess contact number, $\dz$, with strain $\epsilon$ at the jamming transition: $\dz \sim \epsilon^{1/\delta_\e}$, and for the dependence of various quantities on $s$ (e.g. Fig.~10 of Ref.~\cite{Goodrich:2014iu}) when $s$ is controlled as in the fixed-$p s N$ ensemble~\cite{DagoisBohy:2012dh}.

The scaling collapses shown in Figs.~\ref{fig:energy_scaling} and~\ref{fig:s2_scaling} for three-dimensional systems are even more successful if one includes analytic corrections to scaling (see Appendix~\ref{sec:corrections_to_scaling}). Similar collapses are also shown for $d=2$ in Appendix~\ref{sec:corrections_to_scaling}; the collapses are not quite as successful, likely due to logarithmic corrections to scaling expected in the upper critical dimension, as observed previously~\cite{Goodrich:2014iu,vanDeen:2014kl}. We note that in $d=2$, we can convert $N \sim L^2$ to rewrite the argument $N \dz^\psi$ in Eq.~\eqref{eq:scaling_ansatz} as $L \dz^\nu$, where $\nu$ is a correlation length exponent.  We see that the resulting length scale, $\xi \sim \dz^{-\nu}$, has $\nu=\psi/2=1/2$.  This length scale has the same scaling as the crossover length scale for transverse sound, $\ell_T$, which diverges at the jamming transition~\cite{Silbert:2005vw,Schoenholz:2013jv}.

In standard scaling theories, one expects the free energy density to scale as $T/\xi^{d}$ from dimensional analysis, for dimensions at or below the upper critical dimension (hyperscaling).  Eq.~\eqref{eq:scaling_ansatz} would then imply that $\dz^\zeta = \dz^{d \nu}$ so that $\zeta=d \nu$.  In our case, $\nu=1/2$ in $d=2$ and $\zeta=4$, violating hyperscaling.  This violation occurs because the jamming transition lies at $T=0$, so  dimensional analysis fails.

To extend to nonzero temperatures near the jamming transition, it is best to convert to the fixed-$p s N$ ensemble, since strains become problematic in systems that can undergo rearrangement events.  In that case, the scaling ansatz for the free energy becomes 
\eq{	F\left(\dz,p,s,N,T\right) = \dz^\zeta \mathcal{F}_0\left(\frac{p}{\dz^{\beta_p}}, \frac{s}{\dz^{\beta_s}},  N \dz^\psi , \frac{T}{\dz^{\delta_T}} \right). \label{eq:Tscaling_ansatz}}
If one then argues that at high $T$ the free energy should not vanish or diverge at $\dz=0$ and should scale as $T$, one obtains $\delta_T=\zeta=4$, consistent with the scaling of the crossover temperature $T^* \sim \dz^4$ governing whether the system obeys jamming behavior ($T<T^*$) or glassy behavior ($T>T^*$)~\cite{Ikeda:2013gu,Wang:2013gy}.  However, for $T>0$ issues such as time scales and aging become important and we will leave a more thorough exploration for future work.

The scaling ansatz of Eq.~\eqref{eq:scaling_ansatz} does not describe all of the phenomenology associated with the jamming transition.  We have not yet included dynamics, important to capture vibrational properties needed to 
describe longitudinal and transverse phonons and the scaling of the boson peak
frequency, $\omega^*$~\cite{Silbert:2005vw}.  We have not attempted to generalize the theory to densities below the jamming transition, where similar power-law scalings and diverging length scales arise~\cite{During:2012bsa,Lerner:2012jj,Lerner:2012co}.  We have not explored
the implications for the nonlinear responses necessary for describing avalanches~\cite{Lin:2014hz} and shear flow~\cite{Ulrich:2013km}. Finally, as mentioned at the beginning, we have defined $\dphi$ in terms of a system-specific critical density instead of the infinite-system critical density.  We expect that suitable generalizations of our scaling ansatz will be able to capture all of these behaviors.

In summary, we have proposed a scaling ansatz for the jamming transition. The theory, which has three independent exponents, predicts new exponents that we have verified using numerical simulations. The fact that the jamming transition can be described by a scaling ansatz  implies that the jamming transition--like other critical transitions--exhibits emergent scale invariance, and that it should be possible to coarse-grain the system and study the resulting renormalization group flows. The scaling ansatz is therefore a key step towards a complete theoretical description of the jamming transition capable of systematically including friction, non-spherically-symmetric potentials, three-body interactions and other features of the real world, to understand the extent of universality in the mechanical properties of disordered solids.

%

\appendix
\section{Integrating over the $\dz$ distribution\label{sec:dz_distribution}}
The scaling hypothesis implies that the
probability 
distribution $R(\dz | p, \e, N)$ 
for $\dz$
should take the form
\eq{	
	R(\dz | p,\e,N) &= p^{-1/\beta_p} \mathcal{ R}\left(\frac{\dz}{p^{1/\beta_p}},\frac{\e}{p^{\delta_\e / \beta_p}}, Np^{\psi/\beta_p} \right),
}
where we have taken combinations of the natural scaling variables introduced in Eq.~\eqref{eq:scaling_ansatz} so that only the first variable depends on $\dz$. The prefactor $p^{-1/\beta_p}$ normalizes the distribution.
Changing variables to $W = \frac{\dz}{p^{1/\beta_p}}$, $X = \frac{\e}{p^{\delta_\e / \beta_p}}$, and $Y = Np^{\psi/\beta_p}$, we can integrate over the distribution to obtain
\eq{	\avg{W} = \int \mathcal{R}(W,X,Y)W dW. }
Defining $\mathcal{Z}(X,Y) \equiv Y^{1/\psi} \int \mathcal{R}(W,X,Y) W dW$, we can write this as
\eq{	N^{1/\psi} \avg{\dz}= \mathcal{Z}\left(\frac{\e}{p^{\delta_\e / \beta_p}}, Np^{\psi/\beta_p} \right). }
Dropping the $\avg{.}$ notation and inverting $\mathcal{Z}$ with respect to the second argument, we see that
\eq{	p = N^{-\beta_p/\psi} F_1\left(\frac{\e}{\dz^{\delta_\e}}, N \dz^\psi \right). }
Dividing by $\dz^{\beta_p}$, we can write
\eq{	
	\frac{p}{\dz^{\beta_p}} &= \frac{N^{-\beta_p/\psi}  F_1\left(\frac{\e}{\dz^{\delta_\e}}, N \dz^\psi \right)}{\dz^{\beta_p}}\\
	&= (N\dz^\psi)^{-\beta_p/\psi}  F_1\left(\frac{\e}{\dz^{\delta_\e}}, N \dz^\psi \right)\\
	&=  F_2\left(\frac{\e}{\dz^{\delta_\e}}, N \dz^\psi \right).
}


\section{Pressure-shear stress exponent equality \label{sec:sp_exponent}}
Here we derive a key result, resolving the origin of the difference between
the scaling behavior of the bulk and shear moduli at jamming. 
We address this difference by connecting the scaling of the corresponding stresses (pressure and shear stress), and deriving the exponent relation 
\eq{ \beta_s = \beta_p+\psi/2. \label{eq:PvsS_exponent_relation}}
The difference in scaling can be understood by considering the stress tensor, which is microscopically calculated using~\cite{hansen2006theory}
\eq{	\sigma_{\alpha\beta}=-\frac{1}{V} \sum_k b^{(k)} \hat r_\alpha^{(k)} \hat r_\beta^{(k)}. \label{eq:stressdef}}
Here $\alpha$ and $\beta$ index spatial components, $\vec r^{(k)}=r^{(k)} \hat r^{(k)}$ is the vector connecting the centers of two particles along bond $k$, $\hat r^{(k)}$ is a unit vector, $f^{(k)}$ is the magnitude of the force on the bond, and $b^{(k)}=f^{(k)}r^{(k)}$.
Therefore, the pressure (the typical diagonal components) and the shear stress (the typical off-diagonal components) are set by the same residual forces. However, the off-diagonal components add incoherently, introducing an additional system size dependence in the shear stress.

The pressure of an individual system is
\eq{	p=-\frac{1}{d} \Tr \sigma=\frac{1}{Vd} \sum_k b^{(k)} = \frac{N_b}{Vd}\avg{b}_k  \label{eq:pdef}}
where $N_b=NZ/2$ is the number of bonds and $\avg{\cdot}_k$ indicates an average over all bonds.  We will indicate ensemble averages with a $\Lambda$ subscript, so the ensemble average pressure is 
\eq{	\avg{p}_\Lambda = \frac{N_b}{Vd}\avg{b}_{\Lambda}.	\label{eq:pavg_def}} 
Near the jamming transition, the pressure vanishes but $N_b/Vd$ is slowly varying.  We may therefore regard $N_b/Vd$ as constant so that $\avg{b}_\Lambda$ obeys the same scaling as $\avg{p}_\Lambda$ as in Eqs.~\eqref{eq:p_scaling_form} and \eqref{eq:p_scaling_2}.   

For an ensemble where the pressure is not held fixed, the fluctuations in $p$ are described by
\eq{	\delta p^2 \equiv \avg{ \left( p - \avg{p}_\Lambda \right)^2 }_\Lambda. }
Substituting in Eqs.~\eqref{eq:pdef} and \eqref{eq:pavg_def}, this can be written as
\eq{	
\delta p^2 ={}& \frac{N_b}{V^2d^2} \sum_{k} \left[ \avg{b^{(0)}b^{(k)}}_\Lambda - \avg{b}^2_\Lambda\right] \\
 ={}& \frac{N_b}{V^2d^2} \left[ \avg{b^2}_\Lambda - \avg{b}^2_\Lambda \right] \label{eq:deltap2_decomposed}\\
  &+  \frac{N_b}{V^2d^2} \sum_{k \neq 0} \left( \avg{b^{(0)}b^{(k)}}_\Lambda - \avg{b}^2_\Lambda\right). \nonumber 
}

\begin{figure}
	\centering
	\includegraphics[width=\linewidth]{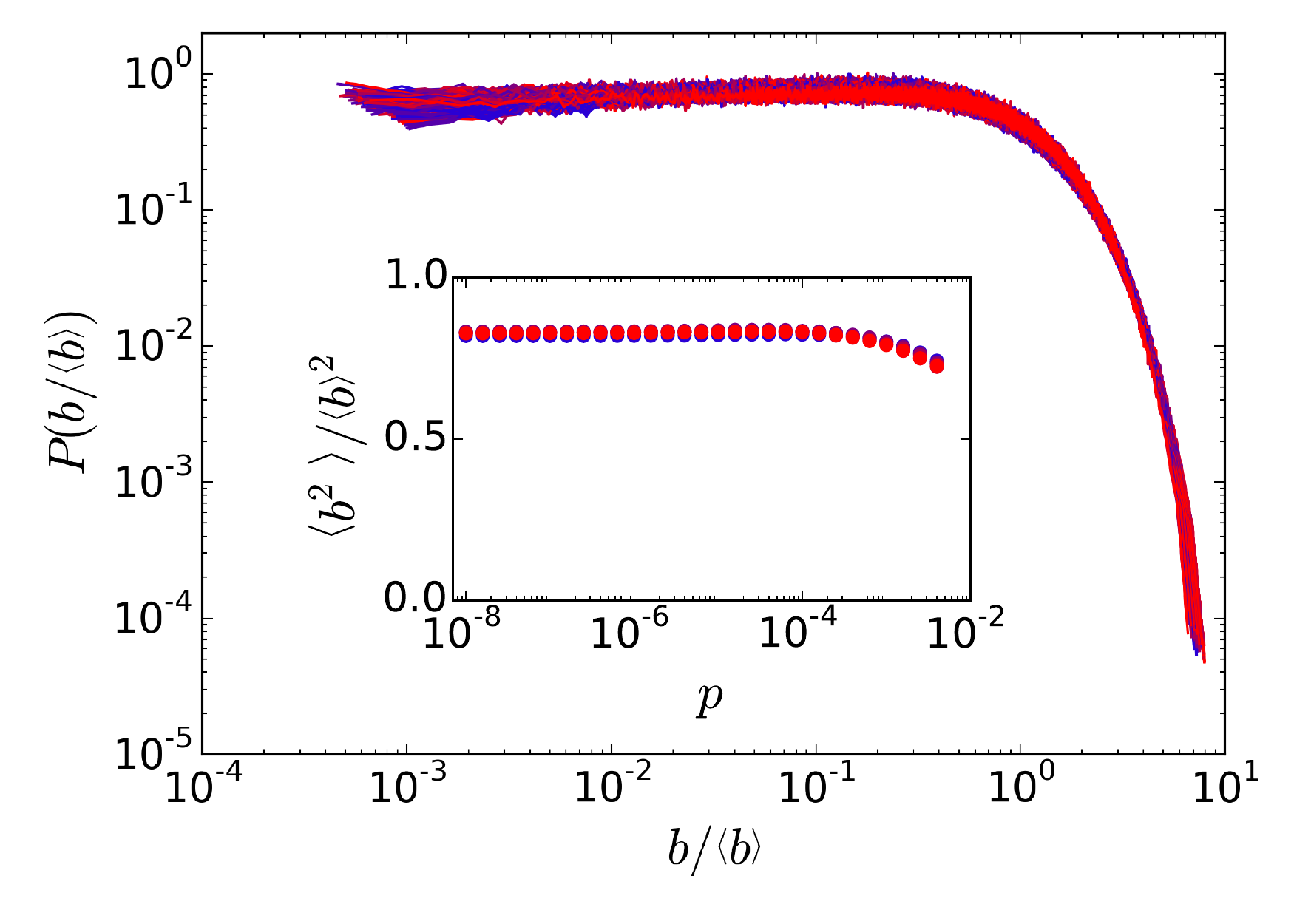}
	\caption{\label{pbdist}Collapse of the distribution $P\left(b/\avg{b}\right)$ for different system sizes and $\dz$. Inset: $\avg{b^2} \sim \avg{b}^2$ over many decades in pressure (or equivalently in $\dz$). Symbols and colors have the same meaning as in Fig.~\ref{fig:energy_scaling}.} 
\end{figure}

We now make two assumptions. First, one would expect that a microscopic quantity like $b^{(k)}$ should have a consistent scaling form, so that the distribution $P\left(b/\avg{b}_\Lambda\right)$ is independent of system size and $\Delta Z$. This is confirmed in Fig.~\ref{pbdist}, and implies that the variance, $\avg{b^2}_\Lambda - \avg{b}^2_\Lambda$, is proportional to $\avg{b}^2_\Lambda$, and therefore scales like $p^2$. 
Similarly, the correlations $\avg{b^{(0)} b^{(k)}}_\Lambda$ with nearby bonds should scale like $p^2$. Second we assume that the correlations between force moments $b^{(k)}$ decays rapidly with the distance between the bonds.
This is consistent with earlier assumptions made to derive the scaling of the shear modulus~\cite{Wyart:2005vu} and with the widespread failure to find long-ranged force correlations in jammed systems.
Thus while there may be 
some very short-range correlations, this should not change the scaling of $\delta p^2$. Finally, since $N_b \sim V \sim N$, we see that 
\eq{	\delta p^2 \sim \frac{p^2}{N}. }
Note that since the pressure is proportional to $\avg{b}_\Lambda$, this result is what one would expect from the central limit theorem.

\begin{figure}
	\centering
	\includegraphics[width=\linewidth]{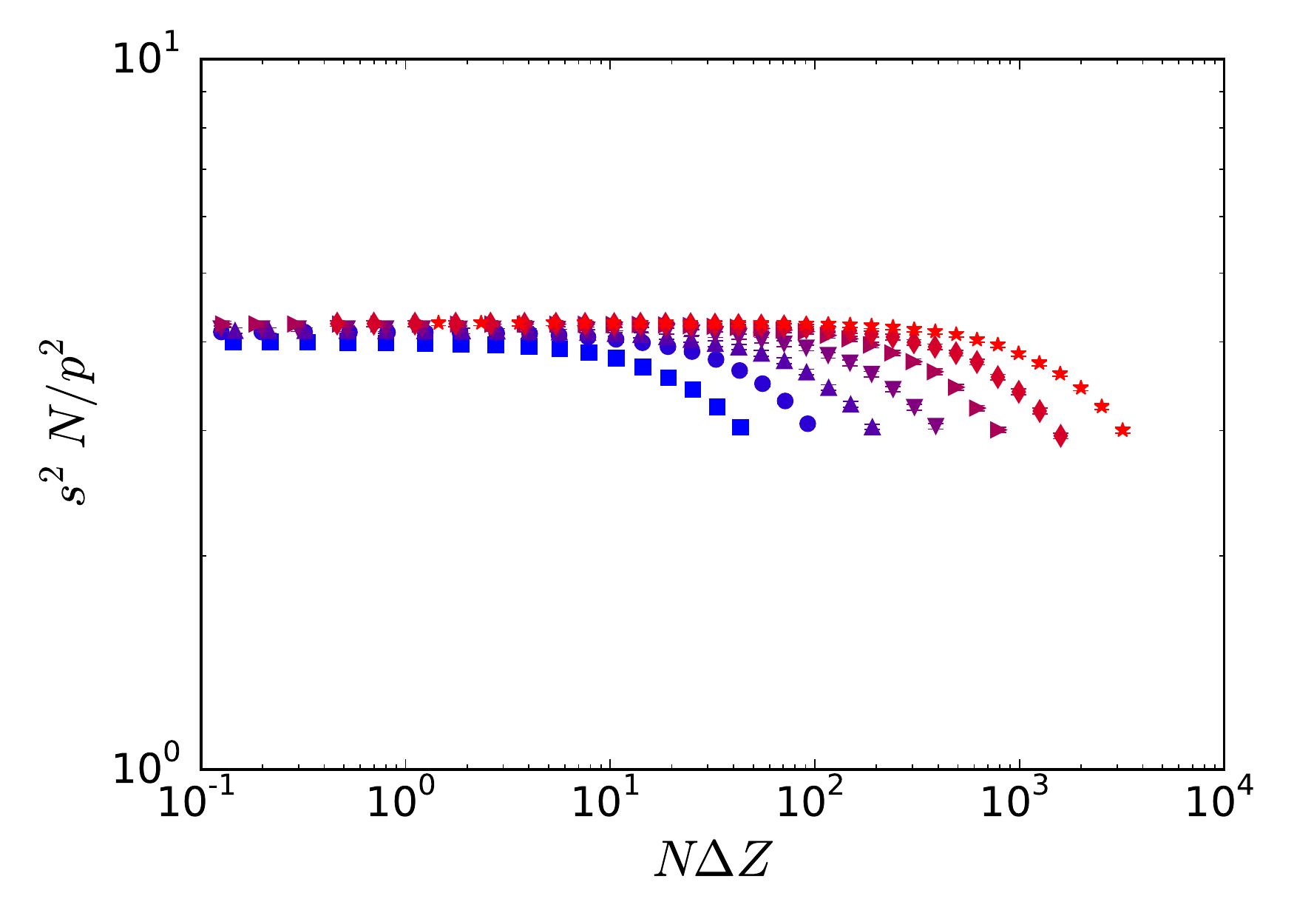}
	\caption{\label{fig:s2N_by_p2_scaling}Verification of the pressure-shear stress exponent relation. $s^2N/p^2$ is constant over several decades in $N\dz$, until $\dz$ is large enough so that analytic corrections to scaling become important (see Appendix~\ref{sec:corrections_to_scaling}).
	Symbols and colors have the same meaning as in Fig.~\ref{fig:energy_scaling}.} 
\end{figure}

We now consider the residual shear stress, which is quantified by the deviatoric stress tensor
\eq{ \tilde \sigma_{\alpha\beta} = \sigma_{\alpha\beta} - \frac 1d \sigma_{\gamma \gamma} \delta_{\alpha\beta}.}
However, $\avg{\tilde \sigma_{\alpha\beta}}_\Lambda =0$ by symmetry, so we instead consider
\eq{	s^2 \equiv \avg{\tilde \sigma_{\alpha\beta} \tilde \sigma_{\alpha\beta}}_\Lambda. \label{eq:s2def}}
Note that this can also be written as $s^2 = \avg{\Tr [\sigma - \frac 1d \Tr \sigma]^2}_\Lambda$.
Substituting Eq.~\eqref{eq:stressdef}, we have
\eq{
	s^2 &=\frac{1}{V^2} \avg{\sum_{kk^\prime} b^{(k)} b^{(k^\prime)} \left[\cos^2(\theta_{kk^\prime}) - \frac{1}{d} \right] }_\Lambda,
}
where $\theta_{kk^\prime}$ is the angle between bonds $k$ and $k^\prime$.  

The sum can again be broken into two pieces:
\eq{
s^2 ={}& \frac{N_b}{V^2} \frac{d-1}{d} \avg{b^2}_\Lambda \label{eq:s2terms} \\
&+\frac{N_b}{V^2} \sum_{k\neq 0} \avg{b^{(0)} b^{(k)} \left [\cos^2(\theta_{0k}) - \frac{1}{d} \right ] }_\Lambda. \nonumber
}
Note that for an isotropic system, $\avg{\cos^2(\theta_{0k}) - \frac{1}{d}}_\Lambda=0$. 
If 
only
short-range correlations exist, then the second term will 
again
have the same scaling as the first term. Thus, as long as there are no long-range correlations, $s^2$ will scale like $\avg{b^2}_\Lambda/N$, and since we
have already seen that $\avg{b^2}_\Lambda \sim \avg{b}^2_\Lambda$, we have 
\eq{	s^2 \sim \frac {p^2}{N}.	}

We therefore expect $s^2 N/p^2 \sim \dz^{2 \beta_s - \psi - 2 \beta_p}$
to be independent of $\dz$ for sufficiently large $N$, implying our
exponent relation $\beta_s = \beta_p+\psi/2$ of
Eq.~\eqref{eq:PvsS_exponent_relation}.
In general, using Eqs.~\eqref{eq:p_scaling_2} and
\eqref{eq:s_scaling_numerical}, we see that 
\begin{equation}
s^2 N/p^2 = f(N\dz^\psi).
\label{s2Np2_scaling}
\end{equation}
Figure~\ref{fig:s2N_by_p2_scaling} shows that this scaling form is obeyed (until analytic corrections become relevant at large $\dz$, see Appendix~\ref{sec:corrections_to_scaling}), and that $s^2 N/p^2$ is constant over several decades of $p$, as expected, affirming our analytical
argument for the scaling relation of Eq.~\eqref{eq:PvsS_exponent_relation}.
Finally, note that while the diagonal and off-diagonal components of the stress tensor scale differently with distance to the critical point, their fluctuations scale the same way.

\begin{figure}[h!tpb]
	\centering
	\includegraphics[width=\linewidth]{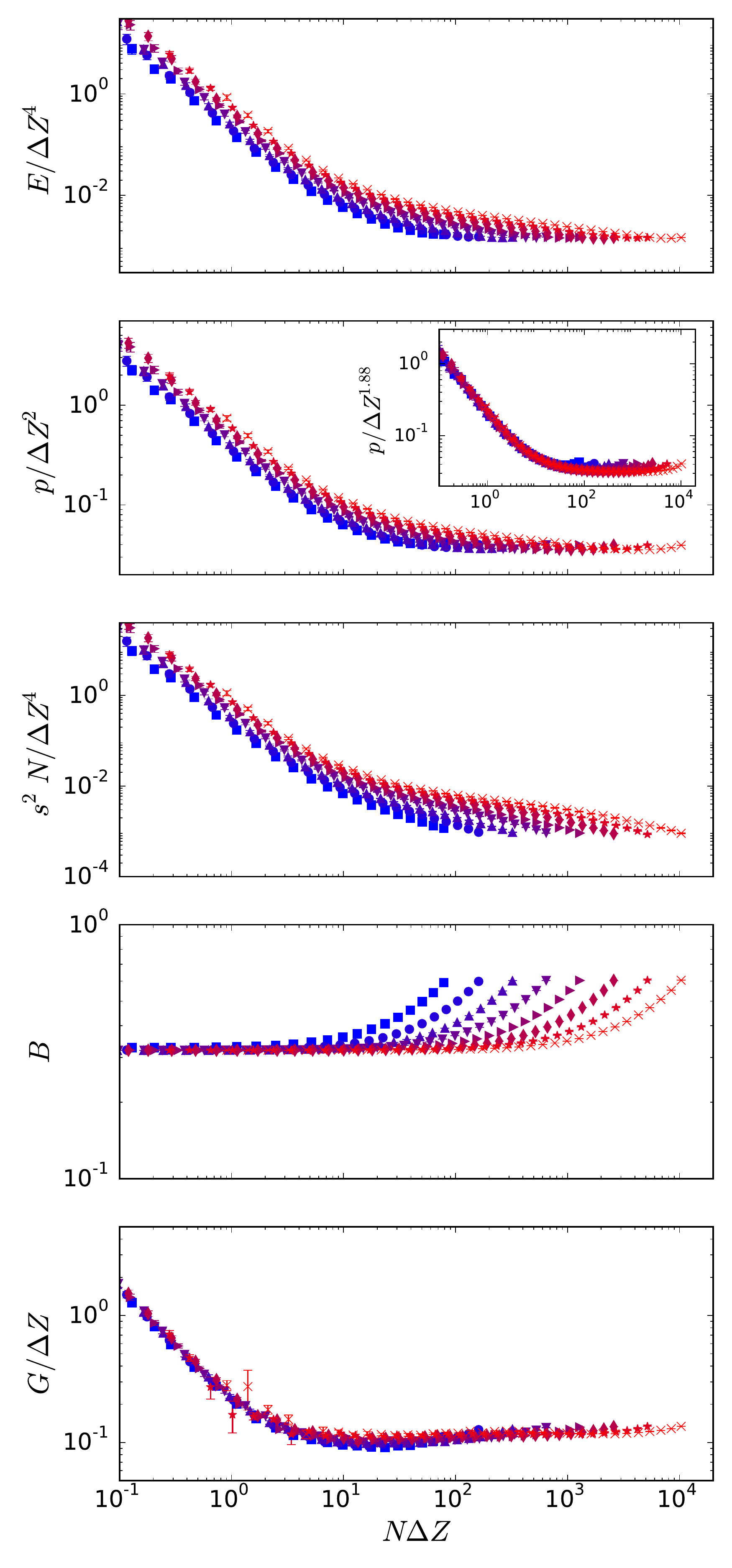}
	\caption{\label{fig:full_scaling_2d}Scaling in 2 dimensions. From top to bottom: the energy, pressure, shear stress, bulk modulus and shear modulus, all scaled according to our theory. Small singular corrections to scaling as well as analytic corrections at large $\dz$ are observed. In $2d$, we include $N=8192$ systems (red crosses); the rest of the symbols and colors have the same meaning as in Fig.~\ref{fig:energy_scaling}.} 
\end{figure}

\begin{figure}[h!tpb]
	\centering
	\includegraphics[width=\linewidth]{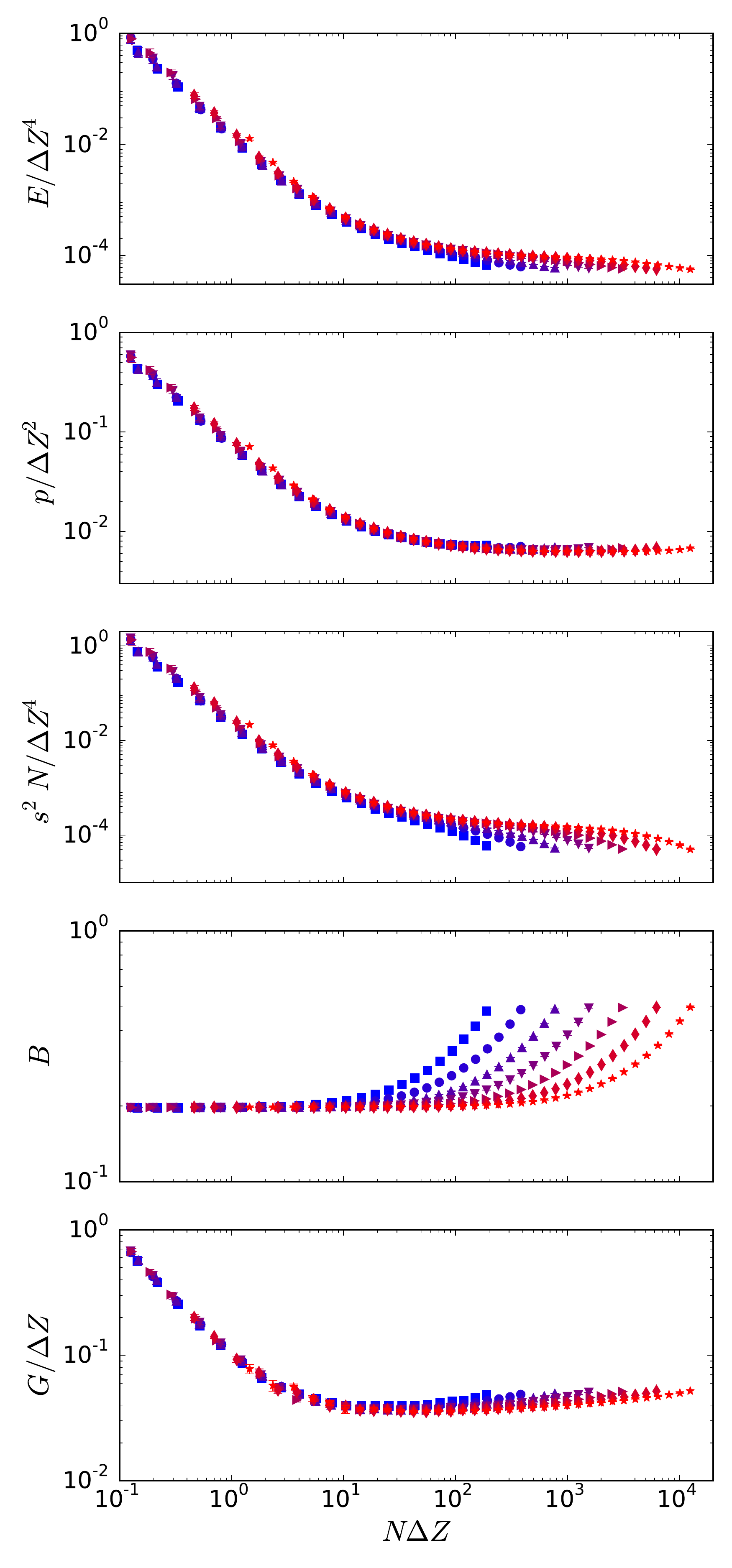}
	\caption{\label{fig:full_scaling_3d}Scaling in 3 dimensions. From top to bottom: the energy, pressure, shear stress, bulk modulus and shear modulus, all scaled according to our theory. Data is extended to higher $\dz$ than what is shown in Figs.~\ref{fig:energy_scaling} and \ref{fig:s2_scaling}, and analytic corrections to scaling are observed in this region. Symbols and colors have the same meaning as in Fig.~\ref{fig:energy_scaling}.} 
\end{figure}

\section{Corrections to scaling\label{sec:corrections_to_scaling}}
In this section, we examine scaling collapses in $d_u = 2$, where we expect singular corrections to scaling to be important.  We also include analytic corrections to scaling at large $\dz$ in $d=3$. Both types of corrections to scaling are expected for critical phase transitions and in no way contradict the ideas presented in the main text. We begin by showing numerical data for the energy, pressure, shear stress, bulk modulus and shear modulus, scaled according to our theory, for systems in the fixed $p\e N$ ensemble in both two dimensions (Fig.~\ref{fig:full_scaling_2d}) and three dimensions (Fig.~\ref{fig:full_scaling_3d}). Note that compared to Figs.~\ref{fig:energy_scaling} and \ref{fig:s2_scaling}, here we extended the data to larger $\dz$. 


The two-dimensional data in Fig.~\ref{fig:full_scaling_2d} show clear systematic deviations from scaling in the energy, pressure, shear stress, and to a lesser extent the shear modulus. Such deviations from scaling are expected at the upper critical dimension of a phase transition, and are not observed in three dimensions (Fig.~\ref{fig:full_scaling_3d}). Also, note that (except for $B$) these deviations are relatively small compared to the singular behavior, which is scaled out in each plot (see Fig.~6 in Ref.~\cite{Goodrich:2014iu}). So, for an extreme example, the energy collapse is good only to a factor of about four, but the energies span a range of 
$10^{14}$
because $\dz$ varies by at least $10^4$. While we do not have a theoretical prediction for the form that such corrections should take, it has been shown that such data can be collapsed by introducing logarithmic corrections~\cite{Goodrich:2014iu,vanDeen:2014kl}. However, we find that including small corrections to the scaling exponents works equally well, {\it e.g.} see the inset to the second plot in Fig.~\ref{fig:full_scaling_2d}.

A scaling ansatz describes only the leading order behavior near a critical point. Subdominant corrections to scaling can, and often, do exist. As one might therefore expect, the scaling collapse in Figs.~\ref{fig:energy_scaling} and \ref{fig:s2_scaling}, as well as the collapse of $B$, $G$, etc., breaks down if the data is extended to larger $\dz$. Figure~\ref{fig:full_scaling_3d} clearly shows this for three-dimensional data where there are no corrections to the critical behavior.

\begin{figure}[h!tbp]
	\centering
	\includegraphics[width=\linewidth]{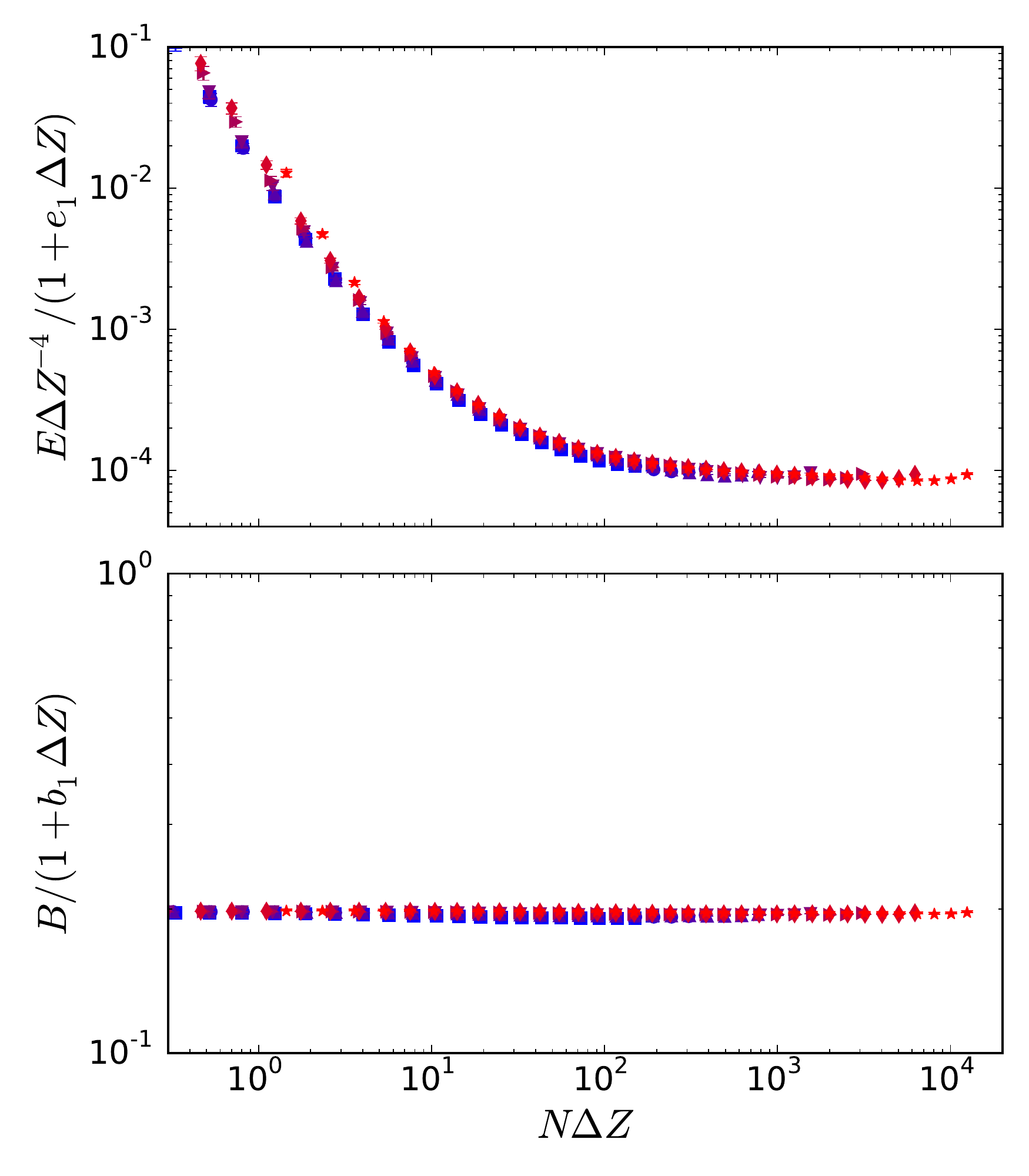}
	\caption{\label{fig:analytic_corrections}Analytic corrections to scaling at large $\dz$ for three-dimensional data. Top: $E \dz^{-4}/\left(1+e_1 \dz\right)$ where $e_1 = -0.13$. Bottom: $B/\left(1+b_1 \dz\right)$ where $b_1 = 0.5$. Symbols and colors have the same meaning as in Fig.~\ref{fig:energy_scaling}.} 
\end{figure}

We find that these deviations at high $\dz$ can be taken into account by incorporating analytic corrections to scaling, obtained by multiplying our scaling ansatz (Eq.~\eqref{eq:scaling_ansatz}) by an analytic function. Expanding this function about the critical point, our ansatz can be written as
\eq{	E = \dz^\zeta \mathcal{E}_0\left(\frac{\dphi}{\dz^{\delta_{\dphi}}}, \frac{\e}{\dz^{\delta_\e}},  N \dz^\psi \right) \left(1 + e_1 \dz + ...\right). \label{eq:scaling_ansatz_analyticcorrections}}
This suggests that the energy should collapse by dividing $E \dz^{-\zeta}$ by $1+e_1\dz$ for some appropriately chosen $e_1$. This is confirmed in the top plot of Fig.~\ref{fig:analytic_corrections}, where we have set $e_1 = -0.13$. 

Analytic corrections are not restricted to the scaling of the energy, and it is clear from our theory how the higher order terms in Eq.~\eqref{eq:scaling_ansatz_analyticcorrections} should propagate to the scaling of other quantities.
For example, including linear corrections to the scaling of the bulk modulus gives
\eq{	B = \dz^{\gamma_B}\mathcal{B}\left( N \dz^\psi \right) \left(1 + b_1 \dz + ...\right), \label{eq:B_scaling_numerical_analyticcorrections}}
suggesting we should get data collapse by dividing $B \dz^{-\gamma_B}$ by $1+b_1 \dz$. This is confirmed by the bottom plot of Fig.~\ref{fig:analytic_corrections} (recall that $\gamma_B = 0$), where $b_1 = 0.5$. Finally, note that in principle, $b_1$ could be a function of the other scaling variables, but we find an excellent collapse can be obtained by keeping it constant.

\begin{acknowledgments}
We thank Sid Nagel and Tom Lubensky for instructive discussions.  This research was
supported by the US Department of Energy, Office of Basic Energy Sciences, Division of
Materials Sciences and Engineering under Award DE-FG02-05ER46199 (AJL, CPG) and by
the National Science Foundation under award DMR 1312160 (JPS). This work was partially supported by a Simons Investigator award from the Simons Foundation to AJL and a University of Pennsylvania SAS Dissertation Fellowship to CPG.
\end{acknowledgments}


%

\end{document}